\documentclass[11pt,a4paper,onecolumn]{article}
\usepackage[utf8]{inputenc}
\usepackage{hyperref}
\usepackage{amsmath}
\usepackage{amsfonts}
\usepackage{amssymb}
\usepackage[capitalize]{cleveref}
\usepackage{graphicx}
\usepackage{color}
\usepackage{appendix}
\usepackage{booktabs}

\hypersetup{
	colorlinks=true,
	linkcolor=black,
	filecolor=black,
	urlcolor=black,
	citecolor=black,
}
\usepackage{transparent}
\usepackage{tikz}
\usepackage[affil-it]{authblk}
\usetikzlibrary{arrows,shapes}
\usepackage[range-phrase = --]{siunitx}
\usepackage[style=phys, sorting=none, backend=bibtex]{biblatex}
\author{Erik Malm\thanks{Electronic address: erikb.malm@gmail.com}}
\affil{{\small MAX IV Laboratory \& Department of Physics, Lund University, 22100 Lund, Sweden}}
\title{\textbf{Phase retrieval via non-rigid image registration}}
\date{}
\begin{document}
	
	\maketitle
	
	\begin{abstract}
	Phase retrieval is the numerical procedure of recovering a complex-valued signal from knowledge about its amplitude and some additional information.
	Here, an indirect registration procedure, based on the large deformation diffeomorphic metric mapping (LDDMM) formalism, is investigated as a phase retrieval method for coherent diffractive imaging.
	The method attempts to find a deformation which transforms an initial, template image to match an unknown target image by comparing the diffraction pattern to the data.
	The exterior calculus framework is used to treat different types of deformations in a unified and coordinate-free way.
	The algorithm performance with respect to measurement noise, image topology, and particular action are explored through numerical examples.
	\end{abstract}

	\section{Introduction} \label{sec:introduction}
	Microscopes that utilize short-wavelength\footnote{Considered as wavelengths within the extreme ultraviolet (\SIrange{5}{50}{\nano \meter}) and X-ray (\SIrange{0.01}{5}{\nano \meter}) regions of the spectrum.} light sources have unique challenges when compared to their visible light-based counterparts.
	They are, in part, due to the difficulty in fabricating efficient, high-numerical aperture optics.
	To circumvent these fabrication challenges microscopes are now being constructed without any image-forming optics, and instead, measure the sample's far-field diffraction pattern directly on a 2D detector.
	In the short-wavelength regime, detectors can only measure a wavefield's intensity which lack any phase information.
	The numerical problem of recovering the phase from these amplitude measurements is commonly referred to as the ``phase problem''.
	Once the phase has been retrieved, the field can be numerically back-propagated from the detector to the sample plane (real-space domain).
	This real-space image can then be related to various sample properties.

	While the phase retrieval problem appears intractable in one dimension,\footnote{A notable exception occurs when sparsity information can be exploited \cite{Sidorenko2015}.} renewed interest in the problem began after numerical results by Gerchberg and Saxton \cite{Gerchberg_Saxton_1972} and Fienup \cite{Fienup1978} showed that the problem could be solved in 2D using alternating projection schemes.
	These simulations were followed by work on uniqueness \cite{Bruck1979,Bates1982,Hayes1982} which suggested that the solution could be expected to be unique\footnote{More precisely, it is a unique equivalence class of solutions related through translation, inversion and global phase rotation.} for dimensions larger than one.
	After these developments, numerous algorithms have been developed to improve the numerical performance \cite{Fienup1982,Elser2003,Luke2004,Szameit2012,Candes2013}.	
	The lack of image-forming optics in these coherent diffractive imaging (CDI) experiments means that the spatial resolution is limited, in principle, by the wavelength of the incoming light and maximum measurement angle, making the method well-suited for imaging experiments which require resolutions in the nanometer range. 
	In addition, CDI is well-suited for studying sample dynamics due to its singleshot capability.
	A drawback to singleshot imaging is the inability to acquire measurements with long exposure times resulting in measurements with relatively low signal-to-noise ratios (SNRs).

	Motivated by previous work on indirect image registration for solving inverse problems \cite{Oktem2017,Chen2018,Gris2020}, this paper explores the use of the large deformation metric mapping (LDDMM) method for phase retrieval.
	The aim of the LDDMM method is to find a diffeomorphism\footnote{A smooth bijective map with smooth inverse.}, $\varphi$, which, through some action, transforms the initial template image, $I_0$, to match the target image, $I_1$. 
	The geometric transformation is given by composition from the right such that $\varphi_. I_0(x) = I_0 \circ \varphi(x)$.
	The mass-preserving action, $\varphi_. I_0(x) = \sqrt{\vert D \varphi (x) \vert} \, I_0 \circ \varphi(x)$, treats the image as a density which preserves the total mass.
	This action has the advantage that the Jacobian determinant,  $\vert D \varphi \vert$, allows the intensity values of the image to be altered, in effect, expanding the number of possible target images which can be recovered.

\begin{figure}[!ht]
		\centering
		{\small
		\def\svgwidth{\linewidth}
		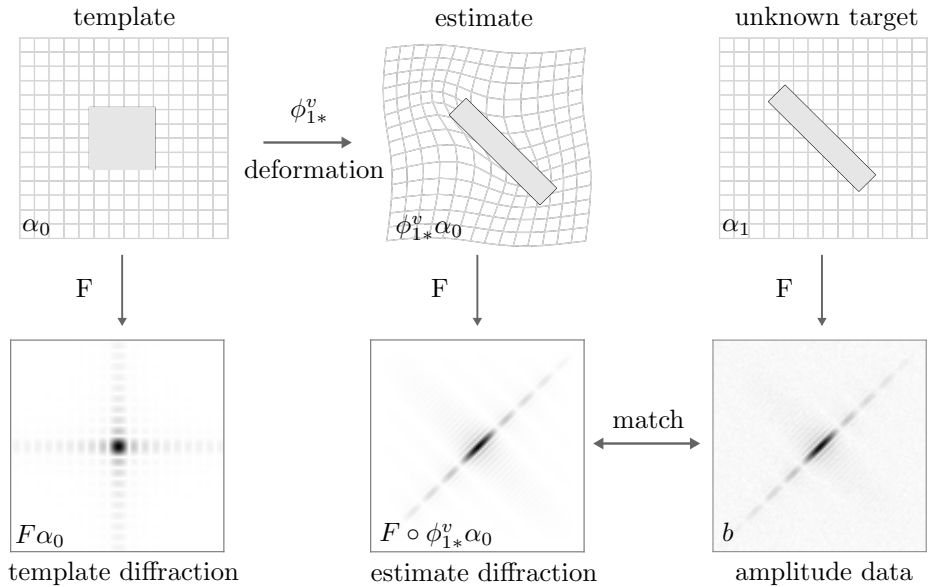
		}
		\caption{Overview of the phase retrieval procedure using indirect image registration. The template image is registered to an unknown target image through $\phi_1^v$ by matching its diffraction pattern to the data, $b$. The nonlinear forward operator, $F$, maps a real-space p-form (image) to far-field diffraction amplitudes.}
		\label{fig:overview}
	\end{figure}

	In our case, we seek a diffeomorphism which transforms the template image until the far-field diffraction pattern matches the data, $b(y)$, $\forall y \in M$, where $M$ is the measurement domain.
	An overview of the method is illustrated in Fig.~\ref{fig:overview}.
	Here, the nonlinear forward operator is given by $F \alpha := \vert \mathcal{F} \alpha \vert$, where $\mathcal{F}$ denotes the 2D Fourier transform, which physically, corresponds to far-field propagation of the sample exit wavefield to the detector plane \cite{Goodman2005,Colton2013}.
	In the numerical simulations we will set the domain to be the flat two-torus, $D = \mathbb{T}^2$.
	Due to the translational invariance of this method, it can be viewed as a registration technique which removes the necessity for an initial, rigid translation to align the template and target images.

	The paper is organized as follows: in Section~\ref{sec:theory}, the theory associated with the LDDMM method is introduced and a derivation of the gradient is given which treats the geometric and mass-preserving actions simultaneously.
	In Section~\ref{sec:numerics}, we define the nonlinear forward operator and consider methods for estimating an appropriate template from the diffraction data.
	The lack of a priori knowledge regarding the sample leads us to consider the cross correlation as the similarity function instead of the usual L$^2$-norm.
	Several numerical examples are provided in Section~\ref{sec:numerical_simulations} which illustrate the algorithm performance under various circumstances.
	Lastly, conclusions and potential extensions to the method are discussed in Section~\ref{sec:conclusion}.

	\section{Theory} \label{sec:theory}
	
	In this section, we will consider the template, $I_0$, and target, $I_1$, images to be real, scalar-valued images in an $n$-dimensional fixed domain, $D$, such that $I_i : D \rightarrow \mathbb{R}$ for $i \in \lbrace 0,1 \rbrace$.
	We seek a diffeomorphism, $\varphi$, which transforms, through some action, $I_0$ to match $I_1$.
	Both the geometric and mass-preserving actions discussed earlier are treated by representing the images as differential forms, $\alpha \in \Omega^p(D)$.
	This approach has been successful in the past in deriving shape derivatives for acoustic and electromagnetic scattering problems by leveraging the framework provided by differential geometry \cite{Hiptmair2013,Hiptmair2018}.
	In component form $\alpha = 1/p! \; \alpha_{i \dots j}(x) \; dx^i \wedge \dots \wedge dx^j$ where $\wedge : \Omega^m(D) \times \Omega^n(D) \rightarrow \Omega^{m+n}(D)$ is the wedge product.
	In this way, the geometric and mass-preserving actions correspond to the push-forward operation, $\varphi_* \alpha$, where $\alpha \in \Omega^0(D)$ or $\alpha \in \Omega^n(D)$ respectively.
	The differential forms are related to the images through the following equations: $\alpha_i = I_i$ or $\alpha_i = I_i \textrm{vol}_g$, where $\textrm{vol}_g$ is the metric volume form in $D$ with metric $g$.
	The push-forward and pull-back operations are related through $\varphi^* \alpha = \varphi^{-1} _* \alpha$ which are given explicitly by $\varphi^* \alpha = I \circ \varphi$ for $\alpha \in \Omega^0(D)$ and  $\varphi^* \alpha = \sqrt{\vert D \varphi \vert} I \circ \varphi$ for $\alpha \in \Omega^n(D)$. Here, $\vert D \varphi \vert$ denotes the Jacobian determinant of $\varphi$.
	Concepts related to exterior calculus will be introduced throughout the text as they are needed, but a more comprehensive treatment can be found in \cite{Fecko2006}.
	
	An exact match of the two images occurs when $\varphi_*\alpha_0 = \alpha_1$.
	In the LDDMM method, $\varphi$ is the endpoint of a curve in the group of diffeomorphisms, $\textrm{Diff}(D)$, which satisfies the flow equation:
	\begin{equation} \label{eq:flow}
	\frac{d \phi^v_t (x)}{dt} = v_t \circ \phi^v_t(x) \, ,
	\end{equation}
	with the initial condition $\phi_0 = \textrm{id}$.
	In this equation, $t \in \lbrack 0, 1 \rbrack$ is a pseudo-time and $v_t(x) \equiv v(t,x)$ is a time-dependent velocity field in the space of smooth vector fields, $V$.

	Next, we develop the form of the energy functional, $\mathcal{E}$, and provide gradient calculations with respect to $v_t$ which will be used to determine $\phi^v_1$.
	The aim is to find the ``smallest'' transformation, $\phi^v_1 \in \textrm{Diff}(D)$, which matches the template and target images.
	The ``smallest'' $\phi^v_1$ is determined using the distance metric in $\textrm{Diff}(D)$ given by
	\begin{equation}\label{eq:distance}
	d(\varphi_0,\varphi_1) = \inf \, \{ \, \int_{0}^{1} \Vert v_t \Vert_V \, dt \, : \, \varphi_1 = \phi^v_1 \circ \varphi_0 \, \} \, .
	\end{equation}
	The norm is given by $\Vert v \Vert_V := \langle Lv, Lv \rangle_{L^2}$. The differential operator, $L$, is defined by $L := -\eta \Delta + \gamma \textrm{id}$, where $\Delta \equiv \nabla^2$ is the Laplacian operator and $\eta, \gamma \in \mathbb{R}$.
	Importantly, the distance metric in \cref{eq:distance} induces a metric between images \cite{Beg2005}.
	This leads us to the following energy functional:
	\begin{equation} \label{eq:cost}
	\mathcal{E}(v_t) := {\underbrace{\sigma \int_{0}^{1} \Vert v_t \Vert^2_{V} dt}_{\mathcal{E}_1(v_t)}}  + { \underbrace{\vphantom{\int_{0}^1 \alpha} \Vert F \circ \phi^v_{1 \, *} \alpha_0 - b \Vert^2_{L^2(M)}}_{\mathcal{E}_2(v_t)}} \, .
	\end{equation}
	A comparison with \cref{eq:distance} shows that $\mathcal{E}_1$ penalizes diffeomorphisms which deviate significantly from $\textrm{id}$.
	In addition, $\mathcal{E}_1$ also ensures, through the $V$-norm, that $v_t$ are smooth such that $\phi^v_1 \in \textrm{Diff}(D)$ \cite{Chen2018}.
	The second term, $\mathcal{E}_2$, is a similarity measure between the diffraction pattern of the estimate and the noisy data, $b$ in the measurement domain, $M$.
	This term ensures consistency with the measured data.
	The relative importance of the two terms is controlled through the scalar $\sigma \in \mathbb{R}_+$.
	The explicit form of the forward operator, $F: \Omega^p(D) \rightarrow L^2(M)$, will be specified in Section~\ref{sec:forward_operator}.
	
	The aim is to find the time-dependent velocity fields, $v_t$, using a gradient descent approach, then, by integrating Eq.~\ref{eq:flow}, to determine $\phi_{t}^{v}$.
	Here, we provide derivations of $\mathcal{E}$ with respect to $v_t$ using the G\^ateaux derivative.
	The G\^ateaux derivative is defined by
	\begin{align} \label{eq:gateaux_definition}
	\partial \mathcal{E}(v;h) & := \frac{d}{d \epsilon} \biggr\rvert_{\epsilon=0} \mathcal{E}(v + \epsilon h) \, .
	\end{align}
	Below, $\partial_h\mathcal{E}(v)$ will be used as shorthand for $\partial \mathcal{E}(v;h)$.
	The G\^ateaux derivative is related to the gradient through
	\begin{align}
	\label{eq:gateaux_definition2}
	\partial \mathcal{E}(v;h) = \int_{0}^{1} \langle \nabla \mathcal{E}, h \rangle_{V} \; dt \, . 
	\end{align}
	Notice that the $V$-gradient is used in \cref{eq:gateaux_definition2} instead of the usual L$^2$ gradient.

	Now we will derive a single expression for $\nabla \mathcal{E}(v)$ which will be specialized to the specific action by specifying the degree of the differential form and converting the expressions into vector notation.
	Through the direct application of \cref{eq:gateaux_definition} to $\mathcal{E}_1(v)$ we immediately obtain
	\begin{equation} \label{eq:gradient_1}
	\nabla \mathcal{E}_1 (v_t) = 2 \sigma \,  v_t \, .
	\end{equation}
	
	Rewriting $\mathcal{E}_2$ in terms of the residual, $R := F \circ \phi^v_{1 \, *} \alpha_0 - b$, the cost function becomes
	\begin{equation}
	\mathcal{E}_2 (v_t) = \Vert R \Vert^2_{_{L^2(M)}} \, .
	\end{equation}
	Taking the G\^ateaux derivative of the above expression we find
	\begin{equation}
	\partial_h \mathcal{E}_2 (v_t) = 2 \langle R,  \partial_h R \rangle_{L^2(M)}
	\end{equation}
	and upon the application of the chain rule the variation of $R$ becomes
	\begin{align}
	\partial_h R &= \partial F (\phi^v_{1 \, *} \alpha_0 \, ; \, \partial_h (\phi^v_{1 \, *} \alpha_0) )\\
	&= \langle \nabla F(\phi^v_{1 \, *} \alpha_0), \partial_h(\phi^v_{1 \, *} \alpha_0)\rangle_{g} \, ,
	\end{align}
	where the inner product between differential forms $\alpha, \beta \in \Omega^p(D)$ is defined by $\langle \alpha, \beta\rangle_{g} := \int_D \alpha \wedge *_{g} \beta $. 
	The subscript on the Hodge operator, $*_g: \Omega^p(D) \rightarrow \Omega^{n-p}(D)$, indicates its dependence on the metric $g$.
	Reversing the order of the inner products we obtain
	\begin{align}
	\partial_h \mathcal{E}_2(v_t) = 2 \langle R', \partial_h (\phi^v_{1 \, *} \alpha_0 ) \rangle_{g}
	\label{eq:gateaux_E2_01}
	\end{align}
	where $R' \in \Omega^p(D)$ is the real-space residual defined by
	\begin{equation} \label{eq:R'}
	R' := \langle R, \nabla F (\phi^v_{1 \, *} \alpha_0) \rangle_{L^2(M)} \, .
	\end{equation}
	
	The final remaining task is to determine an explicit expression for $\partial_h (\phi^v_{\, *} \alpha_0)$.
	A useful identification is made by inspecting the definition for the Lie derivative on $\beta \in \Omega^p(D)$:
	\begin{equation}
	\mathcal{L}_h \beta = \frac{d}{d \epsilon} \Bigr \rvert_{\epsilon=0}  \phi^{\epsilon h  \, *} \beta \, .
	\end{equation}
	Here, we use the notation, $\phi_{r,s} := \phi_s \circ \phi^{-1}_r$, and find
	\begin{align}
	\partial_h \phi^v_{1 \, *} \alpha_0 &= \frac{d}{d \epsilon} \Bigr \rvert _{\epsilon=0} \; \phi^{v +\epsilon h \, * }_{1,0} \alpha_0 \\
	&= - \phi^{v \, * }_{1,0} \mathcal{L}_k \alpha_0 \, .
	\end{align}
	The last equality is a result of $\mathcal{L}_h$ being a natural operator such that $\phi^{v * }_{1,0} \mathcal{L}_{k} = \mathcal{L}_{h} \phi^{v * }_{1,0}$, where $k = \phi^{v}_{1,0 *} h = \int_{0}^{1} \phi^{v}_{t,0 *} h_t dt$ .
	Upon inserting this result into \cref{eq:gateaux_E2_01}, we find the final, coordinate free form for the G\^ateaux derivative
	\begin{align} \label{eq:abstract_gateaux1_E2}
	\partial_h \mathcal{E}_2 (v_t) &= - 2 \langle R', \phi^{v \, *}_{1,0} \int_{0}^{1} \mathcal{L}_{\phi^v_{t,0 *} h_t} \alpha_0 \, dt \rangle_{g}  \\ 
	&= -2  \int_{0}^{1} \langle \phi^{v *}_{t,1} R', \mathcal{L}_{h_t} \phi^{v \, *}_{t,0}\alpha_0 \rangle_{\phi^{v *}_{t,1} g } \, dt \label{eq:abstract_gateaux} \, ,
	\end{align}
	where the last equality is obtained through the application of $\phi^{v *}_{t,1}$.
	Next, \cref{eq:abstract_gateaux} will be converted into vector notation by specifying the degree of the differential forms and making use of Cartan's identity for the Lie derivative, $\mathcal{L}_h = d i_h + i_h d$, where $i_h$ and $d$ denote the inner product with $h$ and the exterior derivative respectively.

	\subsection{Geometric action}

	Consider the case where the images are zero-forms (functions), $\alpha_0 = I_0 \in \Omega^0(D)$. Then, the Lie derivative is equal to $\mathcal{L}_h = i_h d$ and \cref{eq:abstract_gateaux} becomes
	\begin{align}
	\partial_h \mathcal{E}_2 (v_t) =& -2 \int_{0}^{1} \langle \phi^{v *}_{t,1} R', i_h d \, \phi^{v *}_{t,0}\alpha_0 \rangle_{\phi^{v *}_{t,1} g } \; dt \\
	=& -2 \int_{0}^{1} \int_D i_h \lbrace \phi^{v *}_{t,1} R' \times d (\phi^{v *}_{t,0}\alpha_0 ) \rbrace \; \textrm{vol}_{\phi^{v *}_{t,1} g } \; dt \, .
	\end{align}
	Converting this into vector notation we immediately find an expression for the Hilbert space gradient
	\begin{equation} \label{eq:gradient_zero}
	\nabla \mathcal{E}_2 (v_t) = -2 \mathcal{K} \{ R' \circ \phi^v_{t,1} \times \nabla ( I_0 \circ \phi^v_{t,0}) \vert D \phi^v_{t,1} \vert \} \, ,
	\end{equation}
	where $d f$ becomes $\nabla f$ for functions and the Jacobian determinant term, $\vert D \phi^v_{t,1} \vert$, arises from $\textrm{vol}_{\phi^{v *}_{t,1} g } = \vert D \phi^v_{t,1} \vert \, \textrm{vol}_{g }$.
	The reproducing kernel Hilbert space (RKHS) property was used which states that there exists an operator, $\mathcal{K}$, such that $\langle w, h \rangle_{L^2} = \langle \mathcal{K} w , h \rangle_V$.
	As we would expect, when $F = \textrm{id}$, then $R' = R$ and we recover the same expression for the gradient as the direct LDDMM method \cite{Beg2005}.
	Next, the gradient for the mass-preserving action will be determined by treating the images as volume-forms and converting \cref{eq:abstract_gateaux} into vector notation.

	\subsection{Mass-preserving action}
	
	We take a similar approach as before except now $\alpha_i = I_i \textrm{vol}_g \in \Omega^n(D)$ and the Lie derivative becomes $\mathcal{L}_v = d i_v$.
	This results from the property that all volume-forms are closed, or equivalently, $d \alpha = 0$, $\forall \alpha \in \Omega^n(D)$.
	Before converting \cref{eq:abstract_gateaux} into vector notation we need to rearrange terms to obtain an expression with the same form as \cref{eq:gateaux_definition2}.
	The following substitutions are made to make the equations more concise: $\tau^v = 2 \phi^{v \, *}_{t,0} \alpha_0  \in \Omega^n(D) $, $ R'' = \phi^{v \,*}_{t,1} R' \; \in \Omega^n(D) $ with metric $g' = \phi^{v \, *}_{t,1} g $.
	Starting from the integrand in \cref{eq:abstract_gateaux} we have
	\begin{align}
	- \langle \mathcal{L}_{h} \tau^v, \, R'' \rangle_{g'} 
	&= - \langle d i_{h} \tau^v, \, R'' \rangle_{g'} \\
	&= - \langle \tau^v, \, j_{h} \delta_{g'} R'' \rangle_{g'} \\		
	&= \tau^v \wedge i_{h} d *_{g'} R'' \\
	&= \langle h, \, 2 I_0 \circ \phi^v_{t,0} \times  \nabla (R' \circ \phi^v_{t,1}) \, \vert D \phi^v_{t,0} \vert \rangle _{L^2} \, .
	\end{align}
	We used the adjoint relationships: $i_h = (j_h)^\dagger$, where $j_h \beta := \flat_g(h) \wedge \beta \equiv g(h,\cdot) \wedge \beta$ and $\delta = d^\dagger$, where $\delta$ is the codifferential.
	These relationships are valid in our situation because $D = \mathbb{T}^2$ such that $D$ has no boundary, $\partial D = \emptyset $.
	Utilizing the RKHS property again we find that the $V$-gradient in vector notation is
	\begin{equation} \label{eq:gradient_volume}
	\nabla \mathcal{E}_2 (v_t) = 2 \mathcal{K} \{ I_0 \circ \phi^v_{t,0} \times  \nabla (R' \circ \phi^v_{t,1}) \, \vert D \phi^v_{t,0} \vert \} \, .
	\end{equation}
	Equation~\ref{eq:gradient_1} combined with \cref{eq:gradient_zero} or \cref{eq:gradient_volume} are used in a gradient descent scheme to update the velocities fields.

	\subsection{Cross correlation similarity}

	It will become evident in the next section that we lack sufficient information, even for binary images, to create a template image that is within the orbit of the target image.
	So we need to use the cross correlation function as a similarity measure because it is invariant with respect to multiplicative factors and constant offset within the images.
	The cross correlation energy term is defined as:
	\begin{equation} \label{eq:cross_correlation}
	\mathcal{E}_2 (v_t) := \frac{- \, \langle \bar{\beta}^{v}, \bar{b} \rangle_{L^2(M)}^{2}}{2 \, \Vert \bar{\beta}^{v} \Vert_{L^2(M)}^{2}  \;  \Vert \bar{b} \Vert_{L^2(M)}^{2}} = \frac{- A^2}{2 B C} \, ,
	\end{equation}
	where $\beta^{v} := F \circ \phi_{1 \, *}^{v} \alpha_0$ and $\bar{b} := b -\mu_b$ measures the deviation from its average value, $\mu_b$. 
	In the numerical examples in Section~\ref{sec:numerical_simulations} the average will be taken over the entire domain, $M$, but we could equally use a local (windowed) average as well.

	Proceeding in a straightforward way, we find that $\nabla \mathcal{E}_2$ is obtained by replacing $R'$ in \cref{eq:gradient_zero} and \cref{eq:gradient_volume} with the following expression:
	\begin{equation} \label{eq:cross_correlation_residual}
		R' = \frac{A}{B C} \left( \frac{A}{B} \, \bar{\beta}^{\prime \, v} - \bar{b}^{\prime} \right) \, ,
	\end{equation}
	where $\bar{\beta}^{\prime \, v} := \langle \bar{\beta}^{v} , \, \nabla F (\phi_{1 \, *}^v \alpha_0) \rangle_{L^2(M)}$ and similarly for $\bar{b}^{\prime}$.
	The intuitive meaning of this expression will become clear in the next section when we specify an explicit form for $F$.
	Now that we have expressions for $\nabla \mathcal{E}$ we can define the forward operator which will give us a better understanding of $R'$.

	\subsection{Forward operator} \label{sec:forward_operator}
	In the simplest cases, the data measured by a CDI experiment is proportional to the modulus of the Fourier transform of the sample image.
	Mathematically, the forward operator, $F: \Omega^p \left( D \right) \rightarrow L^2 \left( M \right)$, is described by:
	\begin{equation} \label{eq:Forward_operator}
		F \alpha := \vert \mathcal{F} \alpha \vert \, , 
	\end{equation}
	where the Fourier transform of $\alpha$ is denoted by $\mathcal{F} \alpha = \tilde{\alpha}$.
	The Fourier transform, $\mathcal{F}: \Omega^p \left( D \right) \rightarrow L^2 \left( M \right)$, is given by
	\begin{equation} \label{eq:Fourier_transform}
		\mathcal{F} \alpha = \langle \Phi, \alpha \rangle_{g} \, ,
	\end{equation}
	where the kernel for 2D images is defined as 
	\begin{equation} \label{eq:kernel}
		\Phi(x,k) := \exp(-i k \cdot x) \times
		\begin{cases}
			1 \qquad & \textrm{for } p=0\\
			\textrm{vol}_g \qquad & \textrm{for } p=n
		\end{cases} \, .
	\end{equation}
	The inner product is given by $k \cdot x = k_1 x_1 + k_2 x_2$ where $x$ and $k$ are coordinates in $D$ and $M$ respectively.
	
	We need an expression for the gradient of $F \alpha $ in order to calculate $R'$ from \cref{eq:R'}.
	Starting with the Fourier transform, we immediately find the expression $\nabla \mathcal{F} \left( \alpha \right) = \Phi$ and, applying the chain rule, the gradient of $F \alpha$ becomes
	\begin{align}
		\nabla F \left( \alpha \right) =
		\textrm{Re} \;  \frac{\tilde{\alpha}}{\vert \tilde{\alpha } \vert} \, \Phi^{*} \, ,
	\end{align}
	where $\Phi^{*}$ denotes the complex conjugate of $\Phi$ and $\textrm{Re}$ takes the real part of the complex-valued expression.
	
	Looking back at \cref{eq:gradient_zero,eq:gradient_volume} the meaning of $R'$ now becomes clear, it is the inverse Fourier transform of the residual, $R$, which has acquired the phase from our current estimate in Fourier space, $\mathcal{F} \{ \phi_{1 \, *}^{v} \alpha_0 \}$.

	\section{Numerical implementation} \label{sec:numerics}

	In this section, numerical aspects of the algorithm are discussed which play important roles in the method's performance and implementation.
	We start by specifying the procedure for estimating a template from the data and, in the process, see the importance of using the cross-correlation measure.

	\subsection{Template estimation}
	
	Often times the only information about the sample comes from the diffraction data itself.
	In this case, we need a method capable of finding a ``good'' template image.
	In order for the method to recover $\alpha_1$, $\alpha_0$ and $\alpha_1$ must be related through the push-forward action such that $\alpha_1 = \varphi_* \alpha_0$.
	In other words, $\alpha_1$ must be an element of the template orbit:
	\begin{equation} \label{eq:orbit}
	\mathcal{O} = \{ \varphi_{*} \alpha_0 \, : \; \varphi \in \textrm{Diff}(D), \, \alpha_0 \in \Omega^p(D) \} \, .
	\end{equation}
	Therefore, it is critical to the success of the algorithm to find a good estimate $\alpha_0$ to ensure that $\alpha_1 \in \mathcal{O}$.

	Let us start by inspecting the geometric action for the simplest case when $I_1$ is a binary image.
	In this situation, an estimate for the amplitude, $a_0$, is given by the equation
	\begin{equation} \label{eq:amplitude_estimation}
	a_0 = \frac{\int_D \alpha \, \textrm{vol}_g}{\int_D \textrm{supp} \left( \alpha \right) \, \textrm{vol}_g} =: \frac{m}{\vert S \vert} \, ,	
	\end{equation}
	where $m$ is the total mass and $\vert S \vert$ is the image support size.
	Unfortunately, we have access only, through the inverse Fourier transform of $b^2$, to the autocorrelation function.
	The autocorrelation support size, $\vert A \vert$ is related to $\vert S \vert$ through the geometric ratio $G := \frac{\vert A \vert}{\vert S \vert}$ such that the amplitude becomes
	\begin{equation}
	a_0 = \frac{G m}{\vert A \vert} \, .
	\label{eq:amplitude}
	\end{equation}
	For example, rectangular and triangular supports have geometric ratios of \num{4} and \num{6} respectively.
	The geometric ratio is closely related the constraint ratio which relates the number of constraints to the number of unknowns in phase retrieval \cite{Elser2007}.
	The sensitivity of the method with respect to $G$ is illustrated by the reconstructions shown in Fig.~\ref{fig:G_sensativity}.
	The other unknown quantity in \cref{eq:amplitude_estimation} is the total mass, $m$, of the image.
	Determination of $m$ from $b$ is described next for $\alpha \in \Omega^n(D)$.
	
	Now, let us consider the situation where the images are viewed as volume forms, $\alpha \in \Omega^n(D)$.
	To satisfy the constraint that $\alpha_1 \in \mathcal{O}$, $\alpha_1$ must have the same mass as $\alpha_0$ as the push-forward operation preserves this quantity.
	Unlike the geometric action, the total mass can be determined from the data.
	The total mass is determined from the zero-frequency component of the Fourier transform through the relation:
	\begin{align} \label{eq:mass_estimation}
	m = \int_D \alpha =  \mathcal{F} \alpha \biggr\rvert_{k=0} \, ,
	\end{align}
	where $k$ is a coordinate in $M$.
	For real positive images, the total mass of $\alpha_1$ can be determined from the diffraction through the simple relation $m = b\bigr\rvert_{k=0} $.
	
	Due to its invariance with respect to multiplicative factors, the cross correlation function removes the need to estimate $G$ resulting in a more flexible and robust registration method compared to that which is based on the L$^2$ similarity function in \cref{eq:cost}.

	\subsection{Determination of $\phi^v$} \label{sec:phi_updates}
	
	In this section, the update equations for $\phi^v$ will be briefly described.
	More complete discussions on this topic can be found in \cite{Beg2005,Chen2018}.

	After the velocity fields have been updated, the remaining task is to update $\phi_{t,0}$ and $\phi_{t,1}$.
	Integration of the flow equation (\cref{eq:flow}) with respect to time for $\phi_{0,t}^{v}$, and remembering that $\phi_{0,0}^{v} = \textrm{id}$, we find 
	\begin{align} \label{eq:phi0t_update}
	\phi_{0,t}^{v} (x)  = x + \int_{0}^{t} v_{t'} \circ \phi^v_{0,t'}(x) dt' \, .
	\end{align}
	Discretizing with respect to time we have the approximation that
	\begin{align} \label{eq:phi0t_update}
		\phi_{0,t_j}^{v} & \sim \mathrm{id} + \delta t \; \sum_{i=0}^{j-1} v_{t_i} \circ \phi^v_{0,t_i} \\
		& \sim \left( \mathrm{id} + \delta t \cdot v_{t_{j-1}} \right) \circ \phi^v_{0,t_{j-1}} \, .
	\end{align}
	We use the relation $\phi_{t,0} = (\phi_{0,t})^{-1}$ and keep terms up to first order in $\delta t$ to find
	\begin{align} \label{eq:phi_t0_update}
	\phi^{v}_{t_j,0}(x) = \phi_{t_{j-1},0}^{v}(x - \delta t \cdot v_{t_{j}}(x) )
	\end{align}
	for each time step, $t_j$.
	Here, the time interval, $\delta t$, is inversely proportional to the number of time steps.
	A similar approach is taken for $\phi_{t,1}^{v}$ that yields
	\begin{align} \label{eq:phi_t1_update}
	\phi^{v}_{t_j,1}(x) = \phi_{t_{j+1},1}^{v}(x + \delta t \cdot v_{t_{j}}(x) ) \, .
	\end{align}
	It is important to note that, unlike the method in \cite{Avants2008}, numerically we are not guaranteed that $\phi^{v}_{t_j=1,0} \circ \phi^{v}_{t_j=0,1} = \textrm{id}$.
	Issues related to this discrepancy are avoided by calculating $I_0 \circ \phi^v_{t,0}$ using $I_0 \circ \phi^v_{t,0} = (I_0 \circ \phi^v_{t_j=1,0}) \circ \phi^v_{t_j,1}$.

	\subsection{Numerical algorithm}
		
	An overview of the phase retrieval algorithm is shown in Fig.~\ref{fig:diagram}.
	The algorithm starts at the top left by initializing the variables.
	Gradient descent is used to update the velocity fields at each time step, which are then used to update $\phi_{t_j,0}$ and $\phi_{t_j,1}$.
	Next, $\nabla_{v^{k+1}} \mathcal{E}$ and the new step size, $a^{k+1}$, are calculated.
	The step size is set to keep the largest magnitude of the velocity update fixed to a constant value; this helps to maintain the algorithm's progression towards a solution by preventing the velocity updates from becoming very small.
	The cost function is calculated and the algorithm finishes after $k_{\textrm{max}}$ iterations with an estimate of the target: $\phi_{1,0}^{v \; *} \, \alpha_0$.
	The source code is available at \href{https://bitbucket.org/malmy002/phasereg.git}{https://bitbucket.org/malmy002/phasereg.git}.
	\tikzstyle{decision} = [diamond, draw, fill=white, 
	text width=6.0em, text badly centered, node distance=4.0cm, inner sep=0pt]
	\tikzstyle{block} = [rectangle, draw, fill=white, 
	text width=8.0em, text centered, rounded corners, minimum height=4em, node distance=4.0cm]
	\tikzstyle{block2} = [rectangle, draw, fill=gray!10, rounded corners,
	text width=8.0em, text centered,  minimum height=3.0em, node distance=4.0cm]
	\tikzstyle{block3} = [rectangle, draw, fill=gray!10, rounded corners,
	text width=7.5em, text centered,  minimum height=3.0em, node distance=3.0cm]
	\tikzstyle{label} = [circle, fill=white, node distance=1.5cm]
	\tikzstyle{line} = [draw, -latex']
	\tikzstyle{line2} = [draw, -latex', dashed]
	\begin{figure}[!h]
		\centering
		\begin{tikzpicture}
		\node [block2] (init) {\small Initialize: \\ \footnotesize $\nabla_{v^k} \mathcal{E} = 0$, $a^0 = 1$, $\phi_{t_j,0}^{v} = \textrm{id}$, $\phi_{t_j,1}^{v} = \textrm{id}$.};
		\node [block, below of=init,node distance=2.7cm] (velocities) {\small Update $v_{t_j}$: \\ \footnotesize $v^{k+1}_{t_j} = v^{k}_{t_j} - a^k \nabla_{v^k} \mathcal{E}$};
		\node [block, right of=velocities] (phis) {\small Update \\ $\phi_{t_j,0}^{v}$ \& $\phi_{t_j,1}^{v}$: \\ \footnotesize \cref{eq:phi_t0_update,eq:phi_t1_update}.};
		\node [block, right of=phis] (grad) {\small Calculate $\nabla_{v^{k+1}} \mathcal{E}$: \\ \footnotesize \cref{eq:gradient_1,eq:gradient_zero,eq:gradient_zero,eq:R'}};
		\node [block, below of=grad, node distance = 2.75cm] (stepsize) {\small Determine $a^{k+1}$ \\  s.t. \footnotesize $\Vert a^{k+1} \, \nabla_{v^{k+1}} \mathcal{E} \Vert_{\infty} = c$};
		\node [block, left of=stepsize] (cost) {\small Calculate $\mathcal{E}$: \\ \footnotesize \cref{eq:cost,eq:cross_correlation}};
		\node [decision, left of=cost] (convergence) {\footnotesize $k = k_{\textrm{max}} -1$ ?};
		\node [block3, below of=convergence, node distance = 3cm] (finished) {\small Reconstruction: \\ \footnotesize $\phi_{1,0}^{v \; *} \, \alpha_0$ };
		\node [label, below of=convergence, node distance=1.6cm] (dummy_yes) {};
		\node [label, left of=dummy_yes, node distance = 0.45cm] (yes) {\small Yes};
		\node [label, above of=convergence] (dummy_no) {};
		\node [label, left of=dummy_no, node distance = 0.45cm] (no) {\small No};
		\path [line2] (init) -- (velocities);
		\path [line] (velocities) -- (phis);
		\path [line] (phis) -- (grad);
		\path [line] (grad) -- (stepsize);
		\path [line] (stepsize) -- (cost);
		\path [line] (cost) -- (convergence);
		\path [line] (convergence) -- (finished);
		\path [line] (convergence) -- (velocities);
		\end{tikzpicture}
		\caption{\label{fig:diagram} Outline of the phase retrieval algorithm.}
	\end{figure}
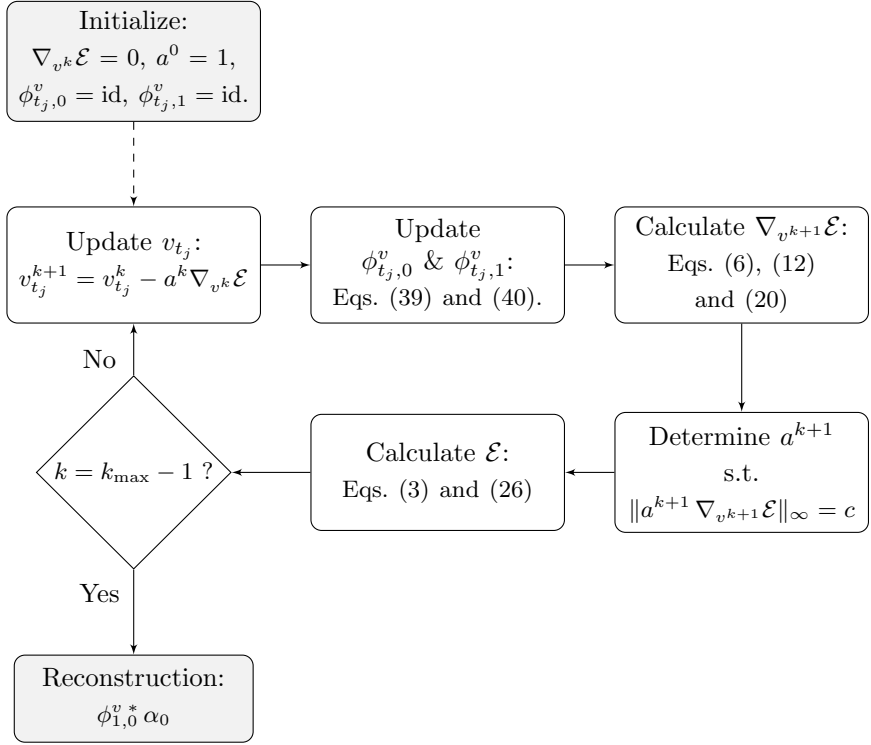

	\section{Simulations} \label{sec:numerical_simulations}
	In this section the algorithm performance in various circumstances is illustrated.
	The results show the sensitivity to the initial template image, consequences of translational invariance of the data fidelity term, $\mathcal{E}_2$, noise robustness, effects of topology within the image and issues related to local minima. 
	As motivation for switching to the cross correlation similarity metric we will look at the reconstruction quality for different $G$ values which are directly related to the amplitude of the template image.
	Next, the performance using the mass-preserving action is shown for two different SNRs.
	The SNR is defined by 
	\begin{align} \label{eq:snr}
		\textrm{SNR} = 10 \log_{10} \left( \frac{\Vert I \Vert_{L^2}}{\Vert I - I_{m} \Vert_{L^2}} \right) \, .
	\end{align}
	In this equation, $I$ and $I_m$ correspond to the ideal and measured (noisy) intensities respectively.
	As the recovered transformations are diffeomorphisms the method is unable to change the topology of the template image; effects related to this are given in the next example which also highlight the importance of inspecting the autocorrelation function before creating the template.
	A comparison between the method and an iterative phase retrieval technique for low-SNR data is given in the final example.
	Results from this simulation show that the method could be useful for imaging dynamic samples using singleshot diffraction data.
	In this case, the template image could be obtained from an initial long-exposure diffraction measurement and used to recover the dynamics within the sample from low-SNR singleshot measurements.

\begin{figure}[!tbp]
	\centering
	\includegraphics[width=1.0\linewidth]{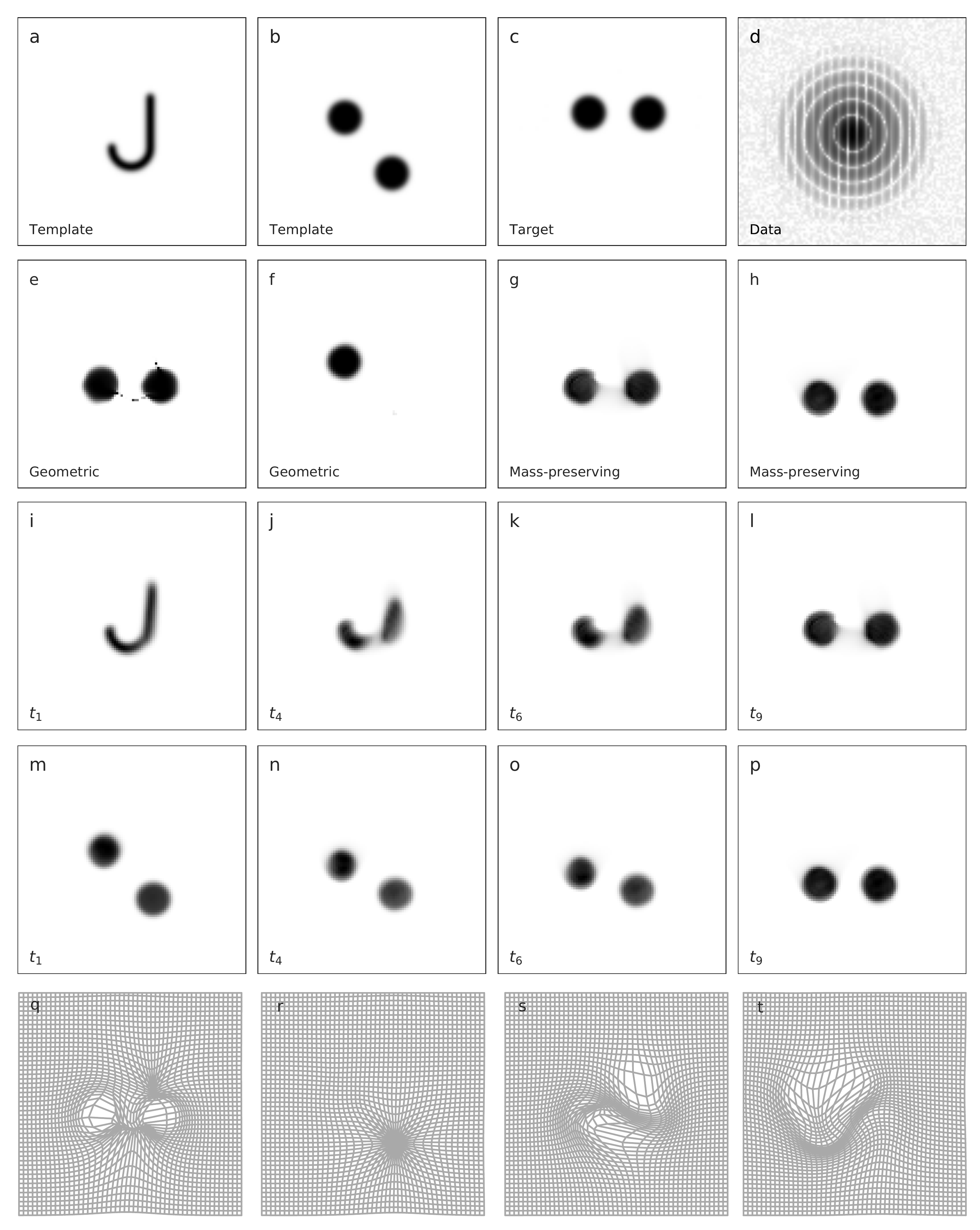} 
	\caption{
		(a,b) Template images with different topologies.
		(c)  The target image and its far field diffraction pattern (d).
		(i--l), (m--p) The reconstruction evolution at different times which used the mass-preserving action and started with the template in (a) and (b) respectively.
		(q,r) The deformations, $\phi^v_{1,0}$, associated with the geometric action for the two different templates.
		(s,t) The deformations, $\phi^v_{1,0}$, associated with the mass-preserving action for the two different templates.
	}
	\label{fig:topology}
	\end{figure}

	For all of the simulations presented, the images have a shape of $100 \times 100$ pixels.
	The relative importance of $\mathcal{E}_1$ and $\mathcal{E}_2$ is fixed by fixing $\sigma=10^{-3}$ and $\gamma=1$ such that $L = - \eta \Delta + \textrm{id}$.
	The time interval, $t \in \lbrack 0 , 1 \rbrack$, is subdivided into \num{10} equal steps such that $\delta t = 1/10$.
	The speed of convergence is increased by fixing $\Vert a^{k} \, \nabla_{v^{k}_{t}} \mathcal{E} \Vert_\infty = 500^{-1}$ for all $t$ with exception to the results shown in Fig~\ref{fig:topology} where it was set to $10^{-3}$.
	The smoothness of the diffeomorphism is controlled through the parameter $\eta$ which was set to $\eta = 5 \times 10^{-3}$ except in the low-SNR simulation (Fig~\ref{fig:low_snr}) where $\eta = 2 \times 10^{-2}$.
	In this case $\eta$ is increased to improve the algorithm's robustness to noise.
	The algorithm is terminated after either \num{1000} or \num{2000} iterations.

\begin{figure}[!tbp] 
	\centering
	\includegraphics[width=1.0\linewidth]{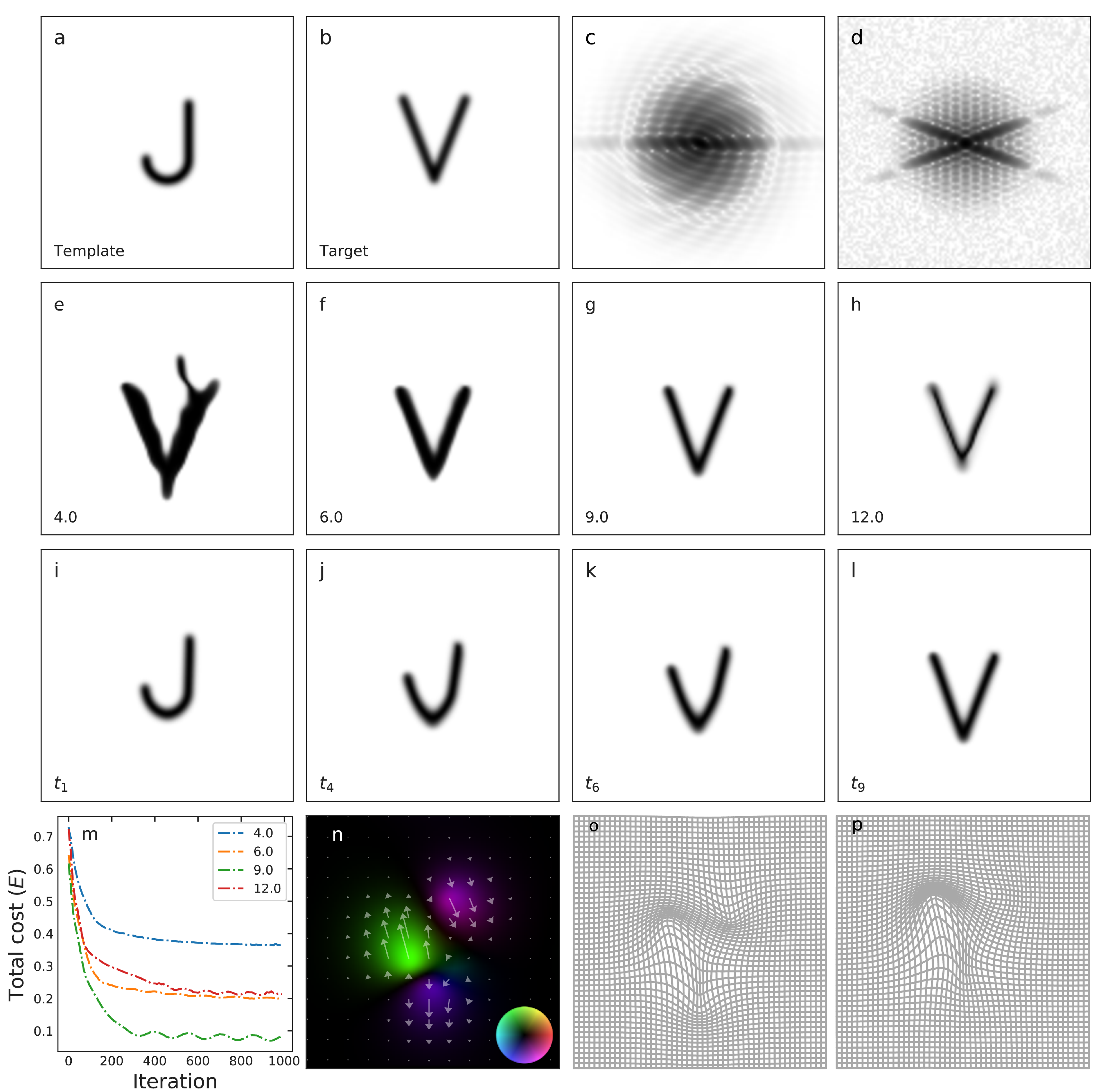} 
	\caption{Algorithm sensitivity to the geometric ratio when $\mathcal{E}_2$ is described by the $L^2$ similarity measure described in \cref{eq:cost}.
	(a,b) The template and target images and their corresponding diffraction amplitudes (c,d) shown in log-scale.
	(d)  The data used by the algorithm to recover the target image.
	(e--h) The final reconstructions for the LDDMM method for different $G$ values which are shown in the bottom left corner of each image. 
	(i--l)  The evolution of the template into the target at different time steps, $t_j$.
	(m) The total energies, $\mathcal{E}$, as a function of iteration number for different $G$ values. 
	(n) The displacement ($\phi_{1,0}^v(x) - x$) associated with the $G=9$ reconstruction.
	The colormap shown in the lower right shows how the hue and value are used to visualize the direction and magnitude respectively. 
	The white arrows are included to aide in this visualization.
	(o,p) Comparison between LDDMM deformations ($\phi^v_{1,0}$) for the indirect and direct methods respectively.
	} 
	\label{fig:G_sensativity}
	\end{figure}

\begin{figure}[!tbp]
		\centering
		\includegraphics[width=1.0\linewidth]{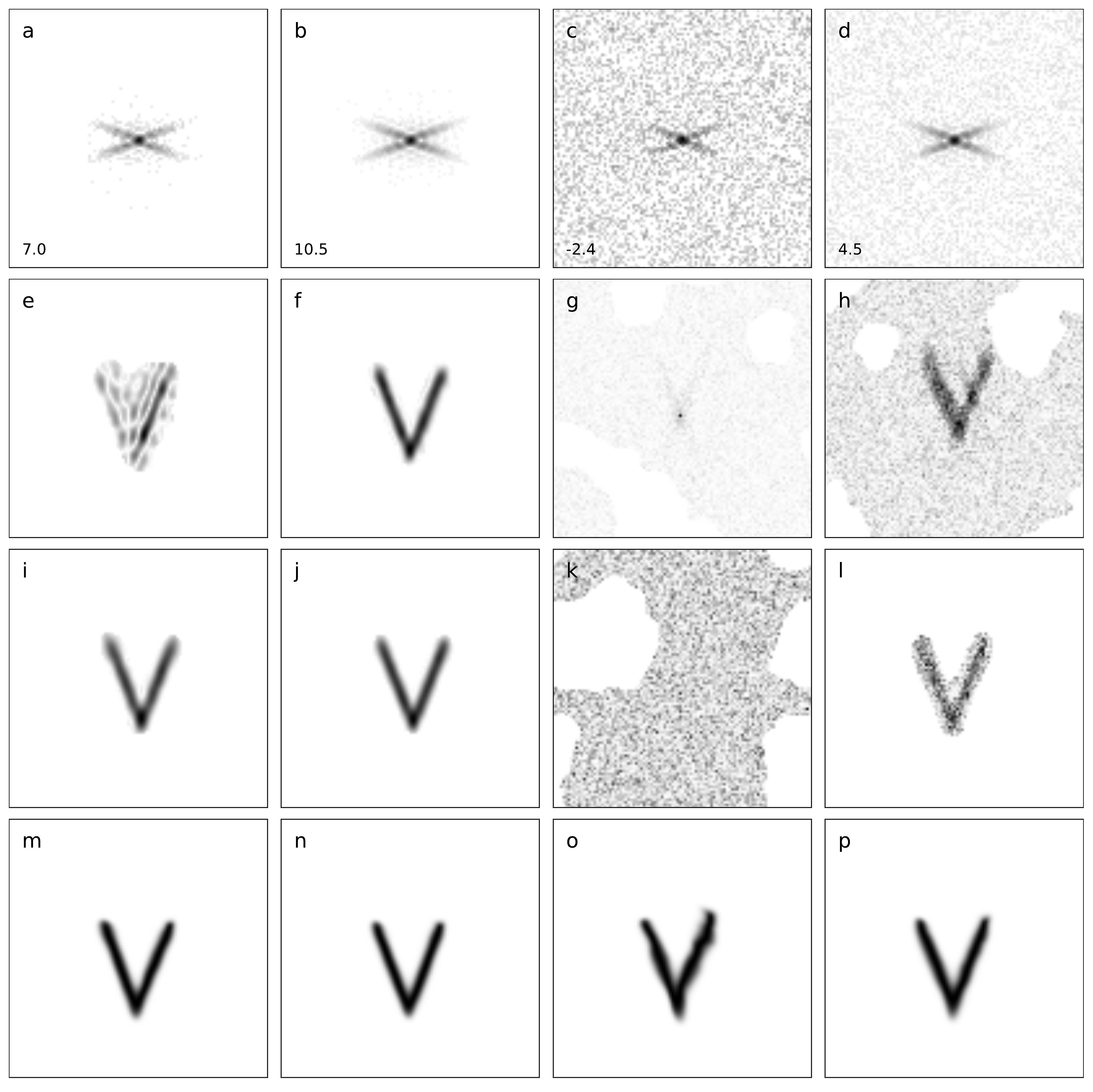} 
		\caption{Comparisons between the proposed algorithm and ER$_{50}$HIO$_{100}$ with shrinkwrap for different amounts and types of noise.
		(a,b) Diffraction data which contains Poisson and quantization noise.
		(c,d) Diffraction data containing Poisson, quantization and Gaussian noise.
		The SNRs are indicated in the lower left corner of the images.
		(e--h) Reconstructions obtained using ER$_{50}$HIO$_{100}$ with a fixed threshold level of \num{0.15} for the shrinkwrap method.
		(i--l) Reconstructions obtained using ER$_{50}$HIO$_{100}$ with different threshold levels for the shrinkwrap method.
		(m--p) Reconstructions using diffeomorphic registration which used the ``J'' image shown in Fig.~\ref{fig:G_sensativity}(a) as the template.
		}
		\label{fig:low_snr}
	\end{figure}

	In the first simulation, we investigate the sensitivity of the method to the geometric ratio described in \cref{eq:amplitude}.
	As $G$ is unknown a priori this will serve as motivation for changing $\mathcal{E}_2$ to a similarity metric based on the cross correlation.
	The reconstructions are provided for $G \in \lbrace 4, 6, 9, 12 \rbrace$.
	We can see from Fig.~\ref{fig:G_sensativity}(e--h) that the method compensates for the smaller image amplitude by increasing the size of the reconstruction.
	The temporal evolution of the method is shown in Fig.~\ref{fig:G_sensativity}(i--l) for $G=9$.
	We can see that the reconstruction appears lower than the target image shown in Fig.~\ref{fig:G_sensativity}(b) which can be expected as $\mathcal{E}_2$ is invariant with respect to global translations.
	This invariance removes the need to find an initial global rigid transformation to align the images and has an interesting effect on the registration procedure.
	Instead of raising the left part of the ``J'' until it matches the target, it raises it only partially while lowering the right side.
	This results in a smaller deformation compared to the direct procedure.
	This can be seen by comparing the indirect (Fig.~\ref{fig:G_sensativity}(o)) and direct (Fig.~\ref{fig:G_sensativity}(p)) deformations.
	The distances associated with the direct and indirect registration methods are found to be \num{0.066} and \num{0.087} respectively which confirms our observation.
	
\begin{figure}[!tbp]
	\centering
	\includegraphics[width=1.0\linewidth]{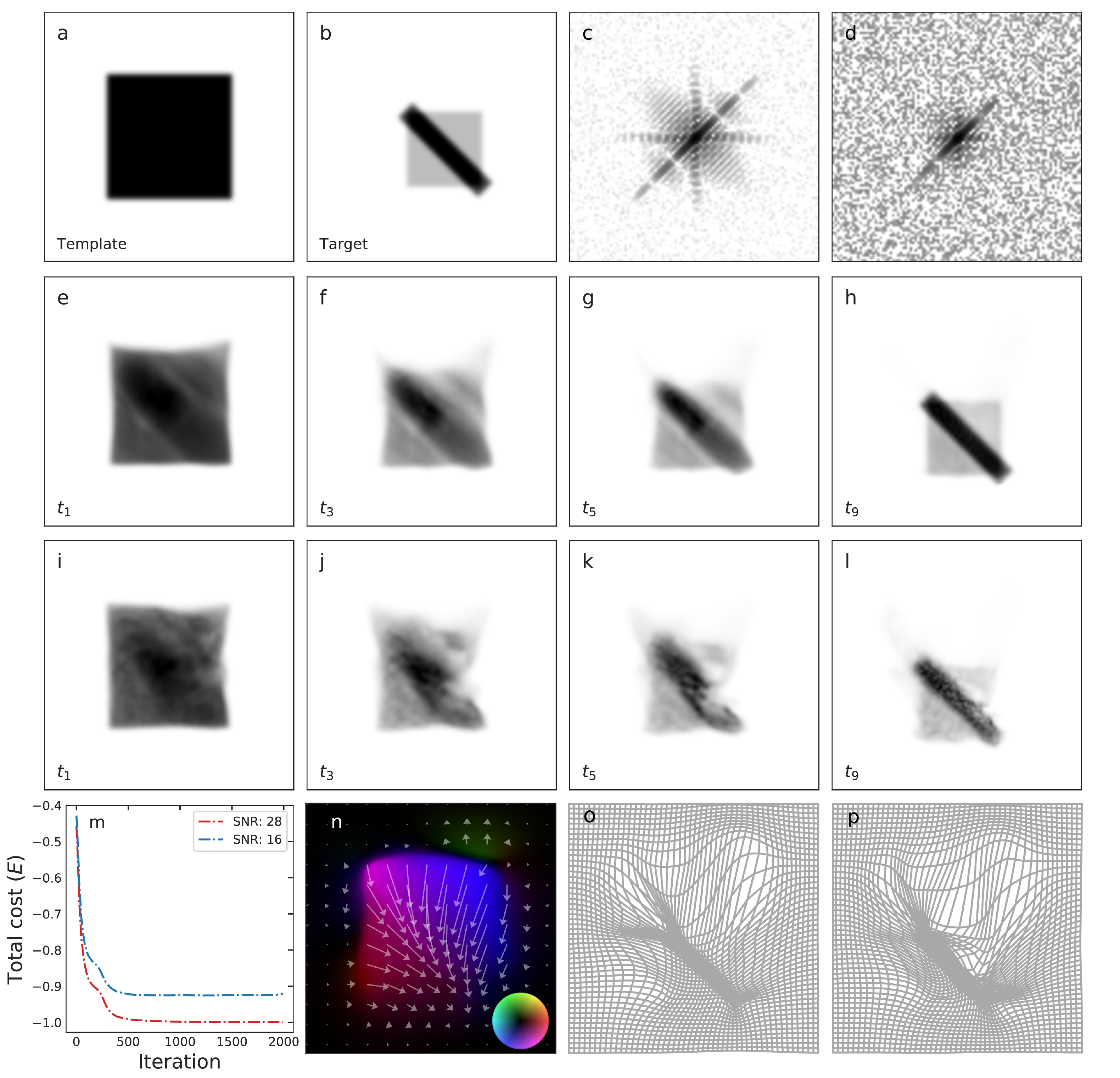} 
	\caption{LDDMM using the volume form action for different SNRs.
		(a,b) The template and target images respectively.
		(c,d) Far-field diffraction amplitudes (log-scale) with SNRs of $28$ and $16$ respectively.
		(e--h) The reconstruction evolution shown at different times associated with the data from (c).
		(i--l) The reconstructions at different times for the data from (d).
		(m) The total energy, $\mathcal{E}$, as a function of iteration number for the two data sets.
		(n) The displacement ($\phi_{1,0}^v(x) - x$) associated with the reconstructions shown in (h).
		The colormap in the lower right illustrates the mapping from direction and magnitude to the image's hue and value. 
		The white arrows are included to aide in this visualization.
		(o,p) Warps, $\phi^v_{1,0}$, associated with the reconstructions using data from (c,d) respectively shown on a coarse grid.
	}
	\label{fig:mass_preserving}
	\end{figure}

	The next simulation illustrates the performance of the algorithm when using the mass-preserving action.
	The template and target images are shown in Fig.~\ref{fig:mass_preserving}(a,b).
	In this case the target image lies outside the orbit of the geometric action and the ability to change the image's intensity values is necessary.
	Figure~\ref{fig:mass_preserving}(c,d) show two diffraction patterns (log scale) with different SNR levels.
	The time evolutions associated with the data in Fig.~\ref{fig:mass_preserving}(c) and Fig.~\ref{fig:mass_preserving}(d) are shown in Fig.~\ref{fig:mass_preserving}(e--h) and Fig.~\ref{fig:mass_preserving}(i--l) respectively.
	The influence of the lower SNR is clearly visible by comparing the final reconstructions.
	The total costs as a function of iteration number are shown in Fig.~\ref{fig:mass_preserving}(m) showing a clear improvement for the higher SNR.
	The displacement is shown in Fig.~\ref{fig:mass_preserving}(n) where the direction and magnitude are mapped to hue and intensity of the image using the color map shown in the inset.
	Lastly the warps associated with the two datasets are visualized in Fig.~\ref{fig:mass_preserving}(o,p) respectively.

	The role of the image topology and the difference in performance between the two actions is highlighted in the next simulation.
	In addition, the importance of using the autocorrelation function to construct the template is shown.
	A comparison between the geometric and mass-preserving LDDMM methods are obtained using the two templates shown in \cref{fig:topology}(a,b) with the target and it's diffraction pattern shown in \cref{fig:topology}(c,d) respectively.
	The reconstructions for the two methods are shown in the second row which clearly shows that the mass-preserving method has obtained superior reconstructions.
	The reconstruction from the geometric method in Fig.~\ref{fig:topology}(f) shows that it was unable to find the correct solution even though the target lies within the orbit of the template.
	This issue can be reconciled by inspecting the autocorrelation function calculated using the data and using a more appropriate template.
	In the density-based method something interesting happens; the method is unable to shrink the ball on the left because the mass must be conserved, so instead, it spreads the mass out until, at some point, it finds the location where the mass should be concentrated.
	In this rare occurrence the method was able to find the global minimum whereas the geometric approach could not.
	The deformations associated with the geometric and density actions are visualized in the bottom row in Fig.~\ref{fig:topology}(q,r) and Fig.~\ref{fig:topology}(s,t) respectively.
	It is surprising that the geometric reconstruction from the template shown in Fig.~\ref{fig:topology}(a) provided a better result than the one from Fig.~\ref{fig:topology}(b) which illustrates the complex dependence of the algorithm on its initial template image and the necessity to inspect the autocorrelation function for constructing the template.

	In the final simulation, we provide a comparison to a method based on an alternating projections strategy for low-SNR data.
	The alternating projection method is based on the error-reduction (ER) and hybrid input-output (HIO) algorithms \cite{Fienup1982}. 
	These algorithms use information about the sample and the diffraction data to find the solution.
	In this case, the sample information is the real-space support ($\{x \in D : I_1(x) \neq 0\}$) which was determined iteratively using the ``shrinkwrap'' method described in \cite{Marchesini2003}.
	The shrinkwrap method is a heuristic approach which determines the support by finding the regions where the smoothed magnitude of the field estimate lies above some threshold value.
	In this case, only the diffraction data and a choice of threshold and smoothing operations are used to determine the solution.
	Diffraction data for four different levels of noise are shown in Fig.~\ref{fig:low_snr}(a-d).
	The SNR is indicated by the number in the lower left corner of the images.
	The data shown in Fig~\ref{fig:low_snr}(a,b) contain only Poisson and quantization noise whereas the data shown in Fig~\ref{fig:low_snr}(c,d) contain additional iid zero-mean Gaussian noise.
	The ideal maximum intensity is set to \num{100} for Fig~\ref{fig:low_snr}(a,c) and \num{500} for the data shown in Fig~\ref{fig:low_snr}(b,d).
	
	In our particular instance, the alternating projection algorithm alternates between \num{50} iterations of ER followed by \num{100} iterations of HIO (ER$_{50}$HIO$_{100}$) which is repeated \num{20} times for a total of \num{3000} iterations.
	This procedure is implemented for \num{20} independent reconstructions.
	The reconstructions were compared to the true target image using the metric described in \cite{Fienup1997,GuizarSicairos2008} and only the reconstructions with the lowest error are shown in Fig.~\ref{fig:low_snr}.
	This is clearly a best-case scenario as in practice the solution would be unknown and we would be unable to compare the reconstructions directly to the solution.
	The images shown in Fig.~\ref{fig:low_snr}(e-h) are the best reconstructions when the shrinkwrap threshold value was fixed to \num{0.15}. 
	Directly below are the best reconstructions for ER$_{50}$HIO$_{100}$ for varying thresholds which are determined heuristically to obtain the best reconstructions.
	The threshold values for the reconstructions in Fig.~\ref{fig:low_snr}(i-l) were set to $\{ 0.2, 0.2, 0.4, 0.3 \}$ respectively.
	Finally, the last row shows the corresponding reconstructions obtained with the geometric LDDMM method after \num{1000} iterations.
	We can clearly see that the ER$_{50}$HIO$_{100}$ method has difficulty with low-SNR data containing Gaussian additive noise.
	On the other hand, our method performed quite well even for very low-SNR data.

	\section{Conclusions} \label{sec:conclusion}
	In conclusion, an alternative approach was developed which treats phase retrieval as an indirect image registration problem. 
	Results from several simulations illustrated the algorithm's sensitivity to parameters such as the geometric ratio, measurement noise and the choice of action.
	The performance was compared to the combined ER-HIO with shrinkwrap method for low-SNR measurements showing the method is potentially well-suited for single-shot coherent imaging experiments.
	
	A number of extensions could be made to this method to improve the numerical performance or expand its application to a larger range of experimental situations.
	Extending the method to include complex-valued images would allow for phase samples.
	The numerical performance could be improved through the implementation of a multi-scale approach similar to the methods described in \cite{Risser2011,Sommer2013}; this would remove some of the need to choose an arbitrary length scale.
	Applying a metamorphosis approach similar to the methods described in \cite{Trouve2005,Gris2020} could help improve the reconstruction quality.
	Adapting the algorithm to avoid getting stuck in local minima is crucial for this method to become a viable alternative to the current state-of-the-art approaches.
	However, we have seen from the numerical examples that this issue can, at least in part, be alleviated by inspecting the autorcorrelation function and choosing an appropriate initial template image.

	\newpage
	
	\clearpage

	\printbibliography

\end{document}